\documentclass[%
 aip,
 jap,
 numerical,
 sd,%
 amsmath,amssymb,
 reprint,%
]{revtex4-1}

\usepackage{graphicx}% Include figure files
\usepackage{dcolumn}% Align table columns on decimal point
\usepackage{bm}% bold math
\usepackage[colorlinks=true,linkcolor=blue,urlcolor=blue,citecolor=blue,breaklinks=true]{hyperref}
\usepackage{mathrsfs}
\usepackage{amsfonts}
\usepackage{braket}
\usepackage{subfig}
\usepackage{textcomp}
\usepackage{breakcites}
\usepackage{textcomp}
\usepackage{pifont}
\usepackage{color}
\newcommand{\cmark}{\ding{51}} %Tick mark in Table III of this manuscript
\newcommand{\xmark}{\ding{55}} % Cross mark in Table III of this manuscript

\newcommand*{\citen}[1]{%
	\begingroup
	\romannumeral-`\x % remove space at the beginning of \setcitestyle
	\setcitestyle{numbers}%
	\cite{#1}%
	\endgroup   
}

\begin{document}

\preprint{AIP/123-QED}

\title[JAP_GaAsNBi_QW]{Electronic bandstructure and optical gain of lattice matched III-V dilute nitride bismide quantum wells for 1.55 $\mu$m optical communication systems}

\author{W. J. Fan}
\affiliation{School of Electrical and Electronic Engineering, Nanyang Technological University, 50 Nanyang Avenue, Singapore 639798, Singapore}
\affiliation{OPTIMUS, Centre for OptoElectronics and Biophotonics, Nanyang Technological University, 50 Nanyang Avenue, Singapore 639798, Singapore}

\author{Sumanta Bose}
\email{sumanta001@e.ntu.edu.sg}
\affiliation{School of Electrical and Electronic Engineering, Nanyang Technological University, 50 Nanyang Avenue, Singapore 639798, Singapore}
\affiliation{OPTIMUS, Centre for OptoElectronics and Biophotonics, Nanyang Technological University, 50 Nanyang Avenue, Singapore 639798, Singapore}

\author{D. H. Zhang}
\email{edhzhang@ntu.edu.sg}
\affiliation{School of Electrical and Electronic Engineering, Nanyang Technological University, 50 Nanyang Avenue, Singapore 639798, Singapore}
\affiliation{OPTIMUS, Centre for OptoElectronics and Biophotonics, Nanyang Technological University, 50 Nanyang Avenue, Singapore 639798, Singapore}

\date{\today}

\begin{abstract}

Dilute nitride bismide GaNBiAs is a potential semiconductor alloy for near- and mid-infrared applications, particularly in 1.55 $\mu$m optical communication systems. Incorporating dilute amounts of Bismuth (Bi) into GaAs reduces the effective bandgap rapidly, while significantly increasing the spin-orbit-splitting energy. Additional incorporation of dilute amounts of Nitrogen (N) helps to attain lattice matching with GaAs, while providing a route for flexible bandgap tuning. Here we present a study of the electronic bandstructure and optical gain of the lattice matched GaN$_x$Bi$_y$As$_{1-x-y}$/GaAs quaternary alloy quantum well (QW) based on the 16-band \textbf{\textit{k$\cdot$p}} model. We have taken into consideration the interactions between the N and Bi impurity states with the host material based on the band anticrossing (BAC) and valence band anticrossing (VBAC) model. The optical gain calculation is based on the density matrix theory. We have considered different lattice matched GaNBiAs QW cases and studied their energy dispersion curves, optical gain spectrum, maximum optical gain and differential gain; and compared their performances based on these factors. The thickness and composition of these QWs were varied in order to keep the emission peak fixed at 1.55 $\mu$m. The well thickness has an effect on the spectral width of the gain curves. On the other hand, a variation in the injection carrier density has different effects on the maximum gain and differential gain of QWs of varying thicknesses. Among the cases studied, we found that the 6.3 nm thick GaN$_3$Bi$_{5.17}$As$_{91.83}$ lattice matched QW was most suited for 1.55 $\mu$m (0.8 eV) GaAs-based photonic applications.

\end{abstract}

%\pacs{Valid PACS appear here.}% PACS, the Physics and Astronomy. Classification Scheme.
%\keywords{Dilute nitride bismide alloy, Electronic bandstructure, \textbf{\textit{k.p}} method, Optical gain}%Use showkeys class option if keyword display desired

\maketitle

\section{\label{sec:intro}Introduction}

The dilute nitride bismide semiconductor, GaNBiAs has attracted much attention recently due to its potential applications in the near- and mid-infrared photonic devices, particularly in the 1.55 $\mu$m GaAs-based laser diodes for fiber optical communication systems \cite{sweeney13,fan13}. Per \% of Bismuth (Bi) incorporation reduces the effective bandgap by $\sim$ 60--80 meV \cite{imhof08}, while Nitrogen (N) helps to achieve lattice matching. The dilute-N-Bi quantum wells (QWs) are therefore expected to have better electron and hole quantum confinement. This fundamental property of the electronic structure makes this material more suitable for telecom laser applications compared to conventional III-V materials such as InP based InGaAsP devices, particularly at high temperature \cite{sweeney11}.

However, to achieve this, a significant count of As atoms must be replaced with Bi atoms. This leads to lattice mismatching and other interface defects. This effect can be compensated for, by an appropriate incorporation of N into the alloy. It has been found that, in GaNBiAs semiconductor alloy, when the N : Bi composition ratio is 0.58 we can obtain lattice matching with GaAs \cite{su14}. Co-alloying N and Bi opens up avenues for bandstructure engineering and precise strain control in the GaNBiAs quaternary alloy \cite{usman13}. In dilute-Bi alloys, the bandgap reduction is caused due to coupling between Bi-resonant state and the valence band (VB) state; while on the other hand, in dilute-N alloys, this effect is induced due to N-resonant state coupling with conduction band (CB) state \cite{song16}. In GaNBiAs alloy, both these effects work independently and simultaneously. The origins of such behavior in dilute-N-Bi alloys can be described using a band anticrossing (BAC) model. Since dilute-N alloys predominantly affect the CB, we use the conduction BAC model. Contrastingly, for dilute-Bi alloys we use the valence BAC (VBAC) model in which the VB is most affected \cite{imhof08}.

In this work, we will study various cases of lattice matched GaNBiAs QWs grown on GaAs barrier, aiming for applications in the 1.55 $\mu$m GaAs-based laser for fiber optical communication systems. A recent work\cite{broderick15} by the O'Reilly group has reported GaBiAs/GaAs QWs, while our work focuses on GaNBiAs QWs, which has better electron confinement due to repulsion between N and the host material's band edge. The larger CB and VB offsets of GaNBiAs/GaAs benefit the QW laser performance, such as rendering a larger characteristic temperature. Nasr \textit{et al.} \cite{nasr15} have worked on GaNBiAs QWs, but limit their investigation only up to optical absorption. Here we shall study the optical gain essential for device design and performance, plus report the modeling and optimization of lattice matched GaNBiAs QW lasers. Gladysiewicz \textit{et al.} \cite{gladysiewicz15} have reported GaInAsBi QWs, which differs from GaNBiAs. To study the electronic structure and optical properties of our GaNBiAs QWs and comment on the performances, we have laid down a \textit{\textbf{k}}$\cdot$\textit{\textbf{p}} 16-band Hamiltonian capable of incorporating the effects of N and Bi doping, as we shall explain now.
 
\section{\label{sec:theory}Theoretical Framework}

Broderick \textit{et al.} had proposed a 12- and 14-band \textit{\textbf{k}}$\cdot$\textit{\textbf{p}} Hamiltonian for GaNBiAs \cite{broderick13}. Here, we propose a 16-band Hamiltonian including the $so_{Bi}$ energy level ($E^0_{Bi,so}$), which was not accounted for in their model. The 16-band model is more accurate, especially for the higher excited states of hole, which can be closer to $E^0_{Bi,so}$, and thus the interaction between them cannot be ignored. We have extended the Kane's 8-band Hamiltonian and used the BAC and VBAC models to form a 16-band Hamiltonian. In addition to Kane's 8-bands, the incorporation of N needs two additional bands to address the local N resonant \textit{s}-like states. On the other hand, Bi introduces \textit{p}-like states, for which six additional bands must be considered with the freedom of spin, including \textit{so} coupling. The 16-band Hamiltonian for GaN$_x$Bi$_y$As$_{1-x-y}$ epilayer, can be written in the form

\begin{equation}
		H_{16\times16}=\left[ \begin{array}{ccc}
		H_{2\times2} & H_{2\times8}  & 0\\
		H_{8\times2} & H_{8\times8} & H_{8\times6} \\
		0 & H_{6\times8} & H_{6\times6} \end{array} \right]
\end{equation}

where $H_{8\times8}$ is the 8-band Hamiltonian given by \cite{fan13}

\begin{widetext}
	\begin{equation}\label{}
	H_{8\times8}=\left[
	\begin{array}{cccccccc}
	
	E^c & c.c. & c.c. & c.c. & c.c. & c.c. & c.c. & c.c. \\
	
	0 & E^c & c.c. & c.c. & c.c. & c.c. & c.c. & c.c. \\
	
	\frac{1}{\sqrt{2}}P_- & 0 & P+Q & c.c. & c.c. & c.c. & c.c. & c.c. \\
	
	-\sqrt{\frac{2}{3}}P_z & \frac{1}{\sqrt{6}}P_- & S^\ast & P-Q & c.c. & c.c. & c.c. & c.c. \\
	
	-\frac{1}{\sqrt{6}}P_+ & -\sqrt{\frac{2}{3}}P_z & -R^\ast & 0 & P-Q & c.c. & c.c. & c.c. \\
	
	0 & -\frac{1}{\sqrt{2}}P_+ & 0 & -R^\ast & -S^\ast & P+Q & c.c. & c.c. \\
	
	-\frac{1}{\sqrt{3}}P_z & -\frac{1}{\sqrt{3}}P_- & \frac{S^\ast}{\sqrt{2}} & -D & -\sqrt{\frac{3}{2}}S & \sqrt{2}R & P-\Delta & c.c. \\
	
	-\frac{1}{\sqrt{3}}P_+ & \frac{1}{\sqrt{3}}P_z & -\sqrt{2}R^\ast & -\sqrt{\frac{3}{2}}S^\ast & D & \frac{S}{\sqrt{2}} & 0 & P-\Delta
	
	\end{array}%
	\right]
	\end{equation}
\end{widetext}

The Hamiltonian contains of both kinetic terms $H_k$ and strain terms $H_\varepsilon$. Detailed expressions of the $H_{8\times8}$ Hamiltonian terms are given in our previous work \cite{fan13}. The $H_{2\times2}$ is the \textit{s}$_\text{N}$-like localized N impurity Hamiltonian and given by \cite{fan13,deng10}

\begin{equation}
	H_{2\times2}=\left[ \begin{array}{cc}
	E_N^0 & 0  \\
	0 & E_N^0  \end{array} \right]
\end{equation}

In a strained III-V-N material, the nitrogen level parameter, $E_N$ is with respect to the valence band maximum (VBM) with strain consideration, such that $E_N$ weakly shifts with applied pressure \cite{vurgaftman03}. However, for the unstrained VBM, the nitrogen level, $E_N^0$ is given by

\begin{equation}
	E_N^0=E_N+VBM^S
\end{equation}

where $VBM^S$ is the top of strained valence band at $\Gamma$-point ($k=0$). This can be calculated by utilizing the strain induced 8-band \textit{H} at $k=0$ point\cite{fan13}.

Similar to N, in order to include the Bi band broadening effect, a modified $H_{6\times6}$ is proposed for the \textit{p}-like localized Bi impurity Hamiltonian \cite{imhof08}, given by

\begin{equation}
	H_{6\times6}=\left[ \begin{array}{cccccc}
	E_{Bi}^0 & 0 & 0 & 0 & 0 & 0 \\
	0 & E_{Bi}^0 & 0 & 0 & 0 & 0 \\
	0 & 0 & E_{Bi}^0 & 0 & 0 & 0 \\
	0 & 0 & 0 & E_{Bi}^0 & 0 & 0 \\
	0 & 0 & 0 & 0 & E_{Bi,so}^0 & 0 \\
	0 & 0 & 0 & 0 & 0 & E_{Bi,so}^0 
	\end{array} \right]
\end{equation}

Similar to N, the Bi levels with respect to the unstrained VBM, $E_{Bi}^0$ and $E_{Bi,so}^0$ are given by

\begin{subequations}
	\begin{equation}\label{}
	E_{Bi}^0=E_{Bi}+VBM^S
	\end{equation}
	\begin{equation}\label{}
	E_{Bi,so}^0=E_{Bi,so}+VBM^S
	\end{equation}
\end{subequations}

The $H_{8\times2}$ Hamiltonian is for the interaction between the \textit{s}$_\text{N}$-like localized N impurity state (with N composition \textit{x}) and host material state and given by

\begin{equation}
	H_{8\times2}=\sqrt{x}V_N\left[ \begin{array}{cc}
	1 & 0\\
	0 & 1\\
	0 & 0\\
	0 & 0\\
	0 & 0\\
	0 & 0\\
	0 & 0\\
	0 & 0
	\end{array} \right]
\end{equation}

Similarly, the $H_{6\times8}$ Hamiltonian is for the interaction between the \textit{p}-like localized Bi impurity state (with Bi composition \textit{y}) and host material state and given by \cite{imhof08}

\begin{equation}
	H_{6\times8}=\sqrt{y}V_{Bi}\left[ \begin{array}{cccccccc}
	0 & 0 & 1 & 0 & 0 & 0 & 0 & 0\\
	0 & 0 & 0 & 1 & 0 & 0 & 0 & 0\\
	0 & 0 & 0 & 0 & 1 & 0 & 0 & 0\\
	0 & 0 & 0 & 0 & 0 & 1 & 0 & 0\\
	0 & 0 & 0 & 0 & 0 & 0 & 1 & 0\\
	0 & 0 & 0 & 0 & 0 & 0 & 0 & 1
	\end{array} \right]
\end{equation}

$V_N$ and $V_{Bi}$ are the coupling coefficient in the BAC model for N and Bi respectively. $H_{8\times6}$ and $H_{2\times8}$ are the transpose of the $H_{6\times8}$ and $H_{8\times2}$, respectively.

We can expand the sixteen dimensional hole envelope wave function for the QWs as

\begin{equation}\label{}
	\Phi _{m}=\left\{ \Phi _{m}^{j}\right\} (j=1,2,...,16)
\end{equation}

where

\begin{equation}\label{}
	\Phi _{m}^{j} = \text{exp}\left[i\left(k_xx+k_yy\right)\right]\sum\limits_ma_{n,m}^j\frac{1}{\sqrt{L}}\text{exp}\left[i\left(k_z+m\frac{2\pi}{L}\right)z\right]
\end{equation}

The QW period is $L=l+d$, where the width of the well is \textit{l} while the barrier width is \textit{d}. $k_i$ $(i=x,y,z)$ are the wavevectors, $a_{n,m}^j$ is the expansion coefficient while \textit{n} is the energy subband index \cite{ng05}.

For the calculation of the optical gain, the squared optical transition matrix element (TME) is important. It is a measure of the momentum of transition strength between the hole and electron subband, and given by \cite{bose16}

\begin{equation}\label{eq:TME}
	\mathcal{P}_{cv,i}=\bra{\Psi_{c,\textbf{\text{k}}}}\textbf{\text{e}}_i\cdot\textbf{\text{p}}\ket{\Psi_{v,\textbf{\text{k}}}}\hspace{0.2cm},\hspace{0.2cm}i=x,y,z
\end{equation}
	
where $\Psi_{c,\textbf{\text{k}}}$ and $\Psi_{v,\textbf{\text{k}}}$ are the real electron and hole wavefunctions respectively, and \textbf{p} is the momentum operator\cite{fan96}. Detailed expressions for the TMEs along the \textit{x}, \textit{y} and \textit{z} directions are furnished in our previous work \cite{fan96}.

A higher average of the TMEs along the \textit{x} and \textit{y} directions would result in a higher transverse electric (TE) mode optical gain, while a higher TME along the \textit{z} direction would lead to a higher transverse magnetic (TM) mode optical gain. The TE mode gain polarized in the \textit{x-y} plane is usually higher for QWs compared to the TM mode gain polarized in the \textit{z} direction.

The linear optical gain spectra is calculated based on the density-matrix theory using the following equations\cite{minch99,ng05}

\begin{equation}\label{eq:gain}
	G\left(E\right)=\left[1-\text{exp}\left(\frac{E-\Delta F}{k_BT}\right)\right]\frac{\pi^2c^2\hbar^3}{n^2E^2}R_{sp}\left(E\right)
\end{equation}

\begin{eqnarray}\label{eq:Rsp}
	R_{sp}\left(E\right)=\frac{ne^2E}{\pi m_0^2\varepsilon_0\hbar^2c^3}\sum\limits_c\sum\limits_v\int\int\frac{\left|\mathcal{P}_{cv}\right|^2}{4\pi^2l}f_cf_v\nonumber\\
	\times\frac{1}{\pi}\frac{\hbar/\tau}{\left(E_{eh}-E\right)^2+\left(\hbar/\tau\right)^2}dk_xdk_y
\end{eqnarray}

where $R_{sp}\left(E\right)$ is the spontaneous emission rate. $f_c$ and $f_v$ are the Fermi-Dirac distributions for the electrons and holes in the CB and VB respectively, and given by \cite{ng05}

\begin{subequations}
	\begin{equation}\label{}
	f_c=\frac{1}{1+\text{exp}\left[\left(E_{en_c}-E_{f_c}\right)/k_BT\right]}
	\end{equation}
	\begin{equation}\label{}
	f_v=\frac{1}{1+\text{exp}\left[\left(E_{hn_v}-E_{f_v}\right)/k_BT\right]}
	\end{equation}
\end{subequations}

where $E_{f_c}$ and $E_{f_v}$ are the electron and hole quasi-Fermi levels respectively, and $\Delta F=E_{f_c}-E_{f_v}$ is the quasi-Fermi energy separation, both of which depend on the injection carrier density. $k_B$ is the Boltzmann constant and \textit{T} is the absolute temperature. Results reported in this paper correspond to 300 K. \textit{e} is electron charge, \textit{E} is photon energy, \textit{n} is refractive index, $\varepsilon_0$ is free-space dielectric constant, \textit{c} is speed of light. $\mathcal{P}_{cv}$ is the TME, $E_{eh}$ is the transition energy and $\tau$ is the intraband relaxation time. Band parameter of GaAs at 300 K are tabulated in Table \ref{tab:GaAs-param}, while the BAC and VBAC parameters to study the effect of N and Bi alloying are tabulated in Table \ref{tab:BAC-VBAC-param}. The references from where these values have been taken are cited alongside. The parameters for the alloy semiconductor are obtained by linear interpolation method.

\begin{table}
	\caption{\label{tab:GaAs-param}Band parameters of GaAs at 300 K}
	\begin{ruledtabular}
		\begin{tabular}{ll}
			\textbf{Parameter (Unit)} & \textbf{Value} \\
			\hline
			\textit{a} (\AA) & 5.65325\footnotemark[1] \\
			$m_e^\ast$ ($m_0$) & 0.067\footnotemark[2] \\
			$\gamma_1$ & 6.8\footnotemark[2] \\
			$\gamma_2$ & 1.9\footnotemark[2] \\
			$\gamma_3$ & 2.73\footnotemark[2] \\
			$E_p$ (eV) & 25.5\footnotemark[1] \\
			$E_{g\Gamma}$ (eV) & 1.424\footnotemark[2] \\
			$\Delta_{so}$ (eV) & 0.341\footnotemark[1] \\
			$a_c$ (eV) & -7.17\footnotemark[3] \\
			$a_v$ (eV) & 1.16\footnotemark[3] \\
			$b$ (eV) & -1.7\footnotemark[3] \\
			$d$ (eV) & -4.55\footnotemark[3] \\
			$C_{11}$ ($10^{11}$ dyne/cm$^2$) & 11.879\footnotemark[2] \\
			$C_{12}$ ($10^{11}$ dyne/cm$^2$) & 5.376\footnotemark[2] \\
		\end{tabular}
	\end{ruledtabular}
	\footnotetext[1]{Ref.~\citen{song16}}
	\footnotetext[2]{Ref.~\citen{Chuang}}
	\footnotetext[3]{Ref.~\citen{chen09}}
\end{table}

\begin{table}
	\caption{\label{tab:BAC-VBAC-param}BAC and VBAC parameters for N and Bi}
	\begin{ruledtabular}
		\begin{tabular}{ll}
			\textbf{Parameter (Unit)} & \textbf{Value} \\
			\hline
			$E_N$ (eV) & 1.65\footnotemark[1] \\
			$V_N$ (eV) & 2.7\footnotemark[1] \\
			$E_{Bi}$ (eV) & -0.4\footnotemark[2] \\
			$E_{Bi,so}$ (eV) & -1.9\footnotemark[2] \\
			$V_{Bi}$ (eV) & 1.55\footnotemark[3]\\
			$\Delta E_{CBM}$ & -2.1\footnotemark[3] \\
			$\Delta E_{VBM}$ & -0.8\footnotemark[3] \\
			$\Delta E_{so}$ & -1.1\footnotemark[3] \\
		\end{tabular}
	\end{ruledtabular}
	\footnotetext[1]{Ref.~\citen{vurgaftman03}}
	\footnotetext[2]{Ref.~\citen{imhof08}}
	\footnotetext[3]{Ref.~\citen{alberi07}}
\end{table}

\section{\label{sec:res-disc}Results and Discussions}

\begin{figure}[h]%[!tbp]
	\centering
	{\includegraphics[width=3.3in]{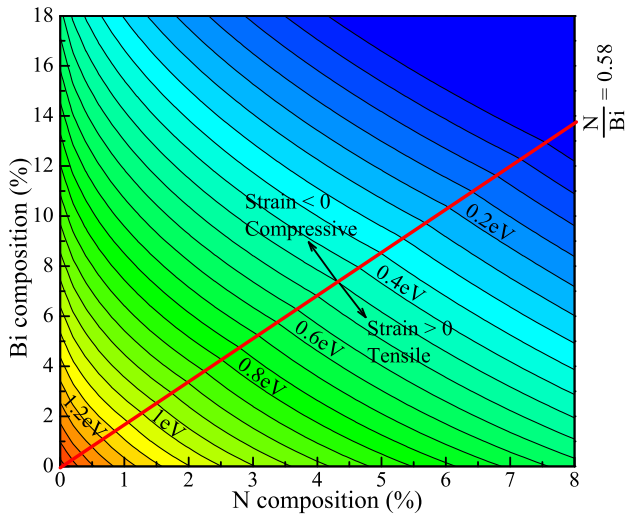}\label{}}
	\caption{Variations in the bandgap ($E_g$) of GaNBiAs pseudomorphically grown on GaAs at 300 K vs. N and Bi composition (\%). The red line is the line of lattice matching (ratio of N : Bi = 0.58).}
	\label{fig:Eg-contour}
\end{figure}

The bandgap (E$_g$) for GaN$_{x}$Bi$_{y}$As$_{1-x-y}$ semiconductor alloy pseudomorphically grown on GaAs depends on the compositions of N (\textit{x}) and Bi (\textit{y}) replacing the As from GaAs. Both GaN and GaBi have lower bandgap compared to GaAs, and combined addition of N and Bi reduces the $E_g$ of GaNBiAs alloy as shown in the Fig. \ref{fig:Eg-contour} contour (This result is at 300 K). The lattice constant of GaN is smaller than GaAs, while that of GaBi is larger. Therefore a sufficiently large N fraction induces a tensile strain ($\varepsilon_{xx}>0$), while a sufficiently large Bi fraction induces a compressive strain ($\varepsilon_{xx}<0$). In this work, we focus on lattice matched GaNBiAs/GaAs QW structures, for which it is essential to maintain a ratio of N : Bi = 0.58 for zero strain, as shown by the red line, which is the line of lattice matching.

\begin{figure}[t]%[!tbp]
	\centering
	{\includegraphics[width=3.3in]{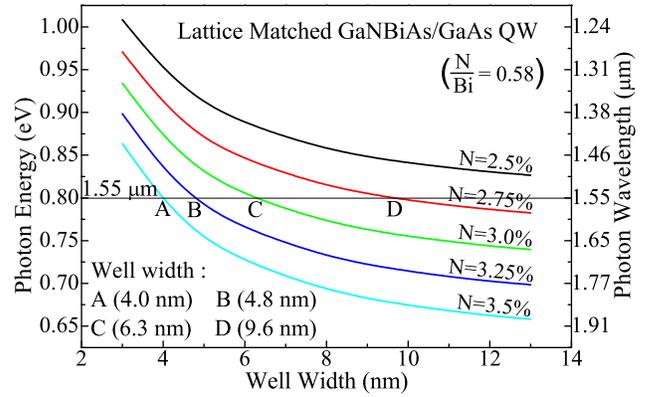}\label{}}
	\caption{Photon emission energy/wavelength vs. well width of lattice matched GaNBiAs QWs with varying N and Bi concentration. Cases A to D are labeled.}
	\label{fig:E-vs-WW}
\end{figure}

From the viewpoint of device physics, strain in the QW material may be beneficial, for example, compressive strain reduces the hole effective mass to improve laser performance. But, from the viewpoint of material growth, strain has the potentially deleterious effects on the performance of laser \cite{rozgonyi73}. Furthermore, the critical thickness of the strained quantum well limits thicker material growth. So, lately there has been many works reporting strain compensated structure to minimize the strain to achieve the high-performance lasers \cite{tansu02,kawaguchi04}. Here, we adhere to the condition of lattice matching, we have studied the photon emission energy of GaNBiAs/GaAs QWs for varying well widths as shown in Fig. \ref{fig:E-vs-WW}. Focusing on the 1.55 $\mu$m emission wavelength optical fiber telecom communication applications, we have chosen 4 cases labeled A to D, with varying well widths of 4, 4.8, 6.3 and 9.6 nm respectively. These cases correspond to varying N compositions of 3.5, 3.25, 3 and 2.75 \% and corresponding Bi compositions for lattice matching with GaAs. The well width increases from Case A to D, and therefore the extent of quantum confinement decreases. Ideally this should lead to a fall in the photon emission energy. But since the composition of N and Bi also reduces simultaneously, this effect is compensated for. A lower fraction of N and Bi is less capable of reducing the effective bandgap of GaNBiAs. Therefore, these two contrasting phenomenon suitably counterbalance and all the four cases exhibit emission around the 0.8 eV (1.55 $\mu$m) mark targeted towards telecom devices, as shown.

\begin{figure*}[t]%[htbp]
	\centering
	\includegraphics[scale=0.61]{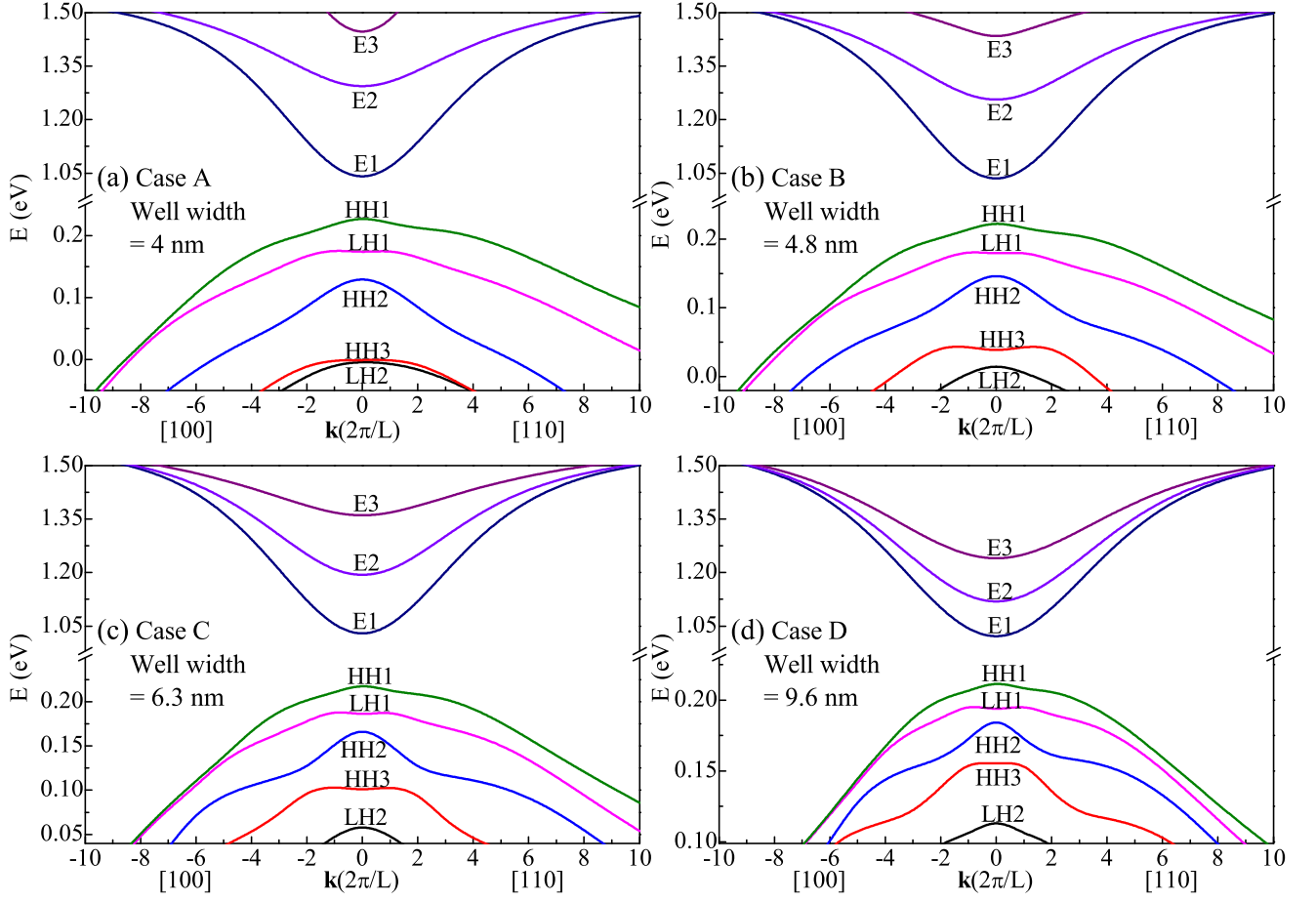}
	\caption{First 3 electron and first 5 hole energy dispersion curves of lattice matched GaNBiAs QWs of (a: \textit{Top Left}) Case A (well width = 4 nm); (b: \textit{Top Right}) Case B (well width = 4.8 nm); (c: \textit{Bottom Left}) Case C (well width = 6.3 nm); (d: \textit{Bottom Right}) Case D (well width = 9.6 nm). \textbf{k} is wavevector along [100] and [110] directions. E stands for electron, HH for heavy hole levels and LH for light hole levels. The photon emission energy for all cases are $\sim0.8$ eV (1.55 $\mu$m).}
	\label{fig:EK}
\end{figure*}

The energy dispersion curves for the first 3 electron and the first 5 holes subband states along the [100] and [110] wavevector direction for Case A to D GaNBiAs/GaAs QWs are shown in Fig \ref{fig:EK}. The well width of each case is mentioned in the figure. The electron states are labeled E1 to E3, while for the holes states we have specified heavy hole (HH) and light hole (LH) states distinctly. We can see that the energy dispersion curves clearly depends on the well width and the (N,Bi) composition. As the well width increases from Case A to D, the (N,Bi) compositions falls, and the two counteracting effects balance to ensure the 1.55 $\mu$m emission. We see that the conduction subband energy dispersion curves are isotropic-like due to the isotropy of electron effective mass. On the other hand, the valence subband energy dispersion curves are more affected by varying well width and composition. There is anisotropy effect in all cases, and they are more spread along the [110] wavevector direction compared to [100] direction. Also, from Case A to D, their \textbf{k} span decreases. But in all four cases, the hole type sequence is same: HH1, LH1, HH2, HH3, LH2. We also notice that as the well width increases from Case A to D, the gap between adjacent E states and adjacent H states reduces. On the whole, $\Delta_{\text{E1-E3}}$ reduces from 403 meV in Case A to 222 meV in Case D. Similarly $\Delta_{\text{HH1-LH2}}$ reduces from 234 meV in Case A to 96 meV in Case D. This can be attributed to the loosening confinement effect. For instance in Case D, due to these effects, E1 and E2 have come very close, and so have HH1 and LH1. This affects the optical performance as we shall subsequently examine.

\begin{figure*}[t]%[htbp]
	\centering
	\includegraphics[scale=0.61]{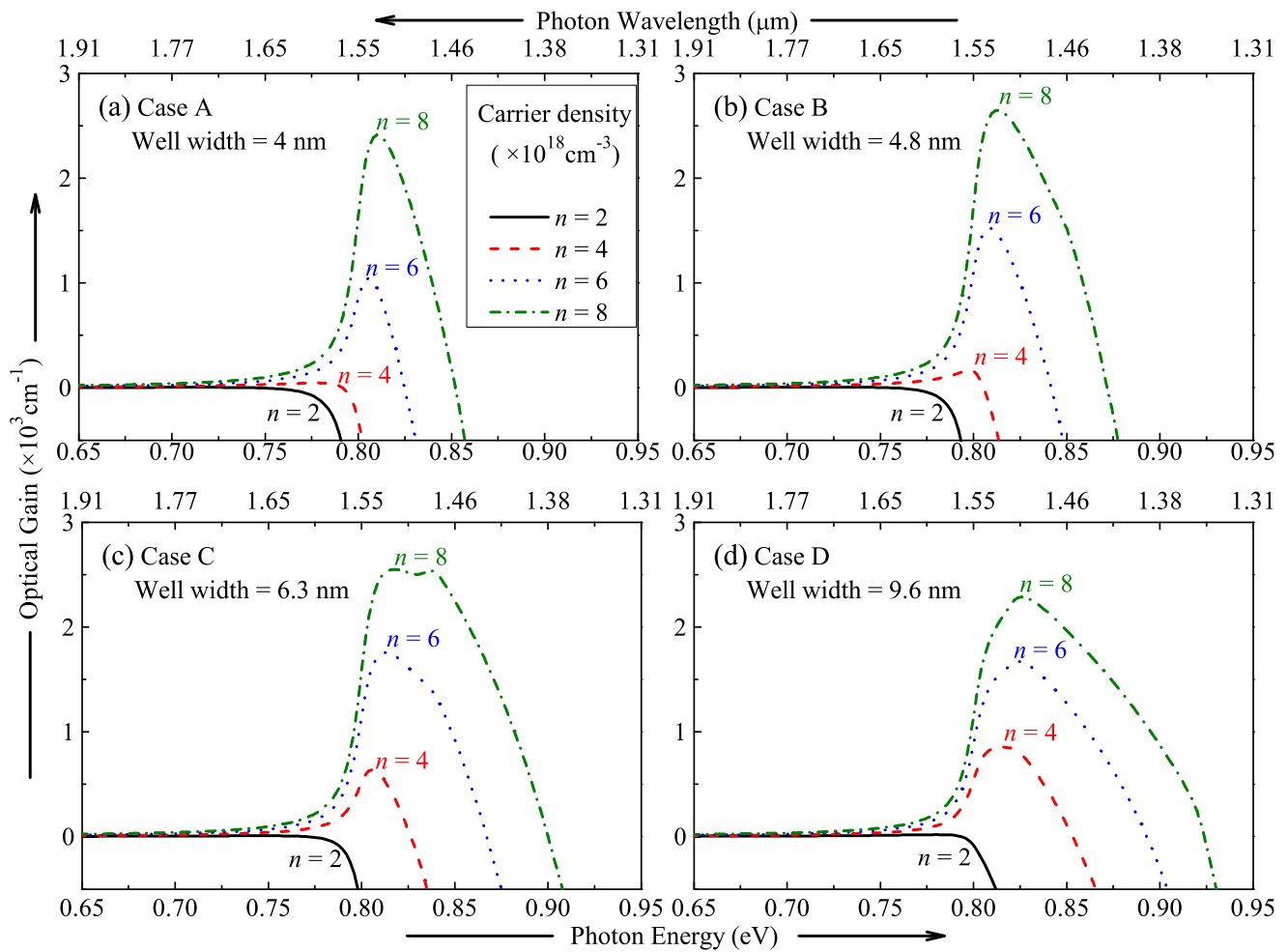}
	\caption{TE mode optical gain spectra of lattice matched GaNBiAs QWs at 300 K for (a: \textit{Top Left}) Case A (well width = 4 nm); (b: \textit{Top Right}) Case B (well width = 4.8 nm); (c: \textit{Bottom Left}) Case C (well width = 6.3 nm); and (d: \textit{Bottom Right}) Case D (well width = 9.6 nm), for varying injection carrier density (\textit{n}) = 2, 4, 6 and 8 $\times10^{18}$ cm$^{-3}$. The peak emission position of all cases are $\sim0.8$ eV (1.55 $\mu$m).}
	\label{fig:wo-mbe-gain}
\end{figure*}

The TE mode optical gain spectra of GaNBiAs QWs of Case A to D are shown in Fig. \ref{fig:wo-mbe-gain}. The injection carrier density was varied from 2 to 8 $\times10^{18}$ cm$^{-3}$. The temperature considered is 300 K. The intraband relaxation time $\tau$ was taken to be 0.1 ps. For any particular GaNBiAs QW case, there is a marginal blue shift in the emission peak position as the injection carrier density increases. With more and more carriers being injected they begin to occupy electronic states further away from the bottom of the conduction band and top of the valence band. Now, the excited state recombinations that incur from such transitions will have their emission energy larger than the E1-HH1 separation, thus resulting in a blue shift. This effect is also generally known as band filling effect in QWs. Comparing Case A to D, we can observe that the band filling effect is more profound in thicker QWs compared to thinner QWs. In thicker QWs, the quantum confinement effect is weaker and therefore the permissible energy levels are much closely packed. With the injection of more and more carriers, there is a higher probability of them occupying the next nearest electronic state and contributing to the emission. Comparing the gain curves, we see that the spectral width increases from Case A to D. There are several factors dictating this trend, the primary being band filling effect as just discussed. Additionally, as seen for Fig. \ref{fig:EK}, the adjacent E and adjacent H states come closer. Under such circumstances if we increase the injection carrier density, there is a high probability that electrons exceed E1 and start to occupy E2 and so on. Similarly holes can exceed HH1 and occupy LH1, HH2, etc. Considering, Case D for instance, it has a fairly large gain spectral width -- because it results from electron-hole recombinations not only from E1-HH1, but also E1-LH1, E2-HH2, etc. However, smaller gain spectral width is better for stable single mode operation.

\begin{figure*}[t]%[htbp]
	\centering
	\includegraphics[scale=0.61 ]{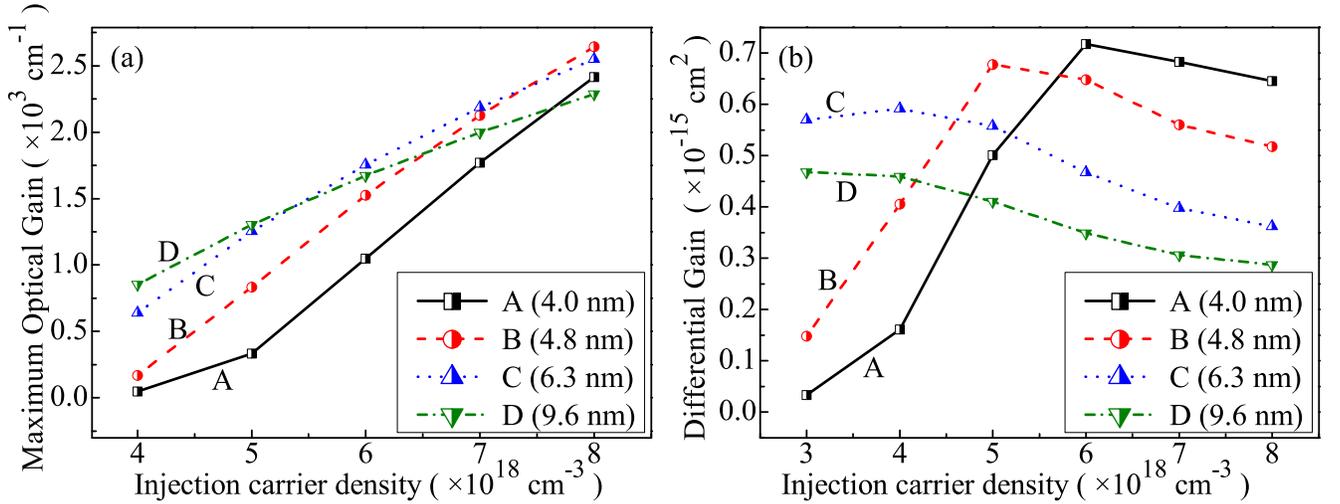}
	\caption{(a: \textit{Left}) Maximum Optical Gain and (b: \textit{Right}) Differential Gain of lattice matched GaNBiAs QWs at 300 K for Case A (well width = 4 nm), Case B (well width = 4.8 nm), Case C (well width = 6.3 nm) and Case D (well width = 9.6 nm) vs. varying injection carrier density. Each case is labeled.}
	\label{fig:wo-mbe-max-diff-gain}
\end{figure*}

The maximum optical gain of the 4 GaNBiAs QW cases studied in Fig. \ref{fig:wo-mbe-gain} are presented in Fig. \ref{fig:wo-mbe-max-diff-gain} (a), with some additional data points. We have ignored 2$\times$10$^{18}$ cm$^{-3}$ because at such low density, there is no positive gain. There are several factors affecting the maximum gain, such as TME, quasi-Fermi levels, temperature and thickness of the QW among others [see factors in Eq. \ref{eq:gain} and \ref{eq:Rsp}]. Usually thicker QWs have higher TMEs due to the higher absolute overlap of the electron and hole wavefunction. Also, the difference between the quasi-Fermi level separation ($\Delta F$) and photon energy (\textit{E}) is important and for positive (maximum) gain, $\Delta F$ must be larger than the fundamental transition energy E1-HH1. Finally, from Eq. \ref{eq:gain} and \ref{eq:Rsp}, we can also see that the gain (thus maximum gain) is inversely proportional to the well width (\textit{l}). Therefore, combining the effects of these and other parameters involved, we get the final result as shown in Fig. \ref{fig:wo-mbe-max-diff-gain} (a). The overall observation is that as the injection carrier density increases, so does the maximum optical gain. This is because a higher carrier density results in more and more electronic states getting filled and greater number of recombinations occur around the \textit{near}-E1-HH1 energy gap. Now comparing the 4 QW cases, we found that their transparency carrier density \cite{quimby-book06-ch23-1} is around the 1.9--3.3$\times$10$^{18}$ cm$^{-3}$ mark. This is the carrier density that gives us neither gain nor absorption around the emission peak position (we get zero gain). But on closer inspection we found that, thinner GaNBiAs QWs have higher transparency carrier density requirement compared to thicker QWs. It was found to be 3.3, 2.5, 2.1, 1.9$\times$10$^{18}$ cm$^{-3}$ for QW Cases A to D respectively, by interpolation method. Just after the onset of gain, say at 4$\times$10$^{18}$ cm$^{-3}$, the maximum gain pattern is Case D\textgreater C\textgreater B\textgreater A, in accordance with their thicknesses. But with further increase in carrier density, the maximum gain is affected by other factors such as band filling effect and reduced confinement. The overall effect on the 4 QW cases can be seen in Fig. \ref{fig:wo-mbe-max-diff-gain} (a). While Case C and D QWs have steadily increasing maximum gain, Case C exceeds D at 8$\times$10$^{18}$ cm$^{-3}$ density, as D approaches saturation. We shall further discuss this effect in the detailed context of Fig. \ref{fig:wo-mbe-max-diff-gain} (b), which shows the differential gain of the 4 GaNBiAs QW cases being studied. A higher differential gain translates to greater modulation speed and narrower width of emission spectra \cite{riane06,anson99}. Thus differential gain is a performance index of how effectively injected carriers produce photon emissions. The thinner QWs (Case A and B) have a steadily increasing differential gain initially, before dipping at sufficiently high densities. On the other hand, the thicker QWs (Case C and D) have a relatively more stable differential gain, but they decrease with increasing carrier densities. Among these, Case D has a lower differential gain and also it is monotonically decreasing from density 2 to 8 $\times$10$^{18}$ cm$^{-3}$. Case C has a higher and more stable differential gain across varying carrier densities.

Now we are in a situation to compare the 4 GaNBiAs QW cases studied, on the basis of their optical gain spectral width [Fig. \ref{fig:wo-mbe-gain}], transparency carrier density, maximum optical gain [Fig. \ref{fig:wo-mbe-max-diff-gain} (a)] and differential gain [Fig. \ref{fig:wo-mbe-max-diff-gain} (b)]. The gain spectral width increases from Case A to D. While Cases A, B and C have a steady form function, for Case D it get very wide with increasing carrier density. This is because at such large thickness, the confinement effect drops and the gap between adjacent E and H states shrinks. Other excited electron-hole transition routes open broadening the spectra. This has an adverse effect on the optical performance, as smaller gain spectral width is better for stable single mode operation. When it comes to the maximum optical gain, Cases C and D have steady increment. Also at sufficiently high carrier density, the Case C maximum gain exceeds that of Case D. The maximum gain of Case B also increases rapidly, but it is much less at lower densities. Moreover, the transparency carrier density of Case C and D are relatively lower than that for A and B. Finally, the study of differential gain shows us that Case C has the best suitable characteristics to deliver high and steady optical performance over a wide range of injection carrier density. The other cases exhibit diminishing differential gain at either low density or high density. These observations are symbolically summarized in Table \ref{tab:comparison}. 

\begin{table*}[t]
	\caption{\label{tab:comparison}Comparative analysis of the 4 GaNBiAs QW cases studied}
	\begin{ruledtabular}
		\begin{tabular}{cccccc}
			GaNBiAs QW Case & N : Bi : As & Gain spectral width & Maximum gain & Transparency carrier density & Differential gain \\
			\hline
			Case A (4.0 nm) & 3.50 : 6.03 : 90.47 & \cmark & \xmark & \xmark & \xmark \\
			Case B (4.8 nm) & 3.25 : 5.60 : 91.15 & \cmark & \xmark & \xmark & \xmark \\
			Case C (6.3 nm) & 3.00 : 5.17 : 91.83 & \cmark & \cmark & \cmark & \cmark \\
			Case D (9.6 nm) & 2.75 : 4.74 : 92.51 & \xmark & \cmark & \cmark & \xmark \\
		\end{tabular}
	\end{ruledtabular}
\end{table*}

\begin{figure*}[t]%[!tbp]
	\centering
	{\includegraphics[scale=0.61]{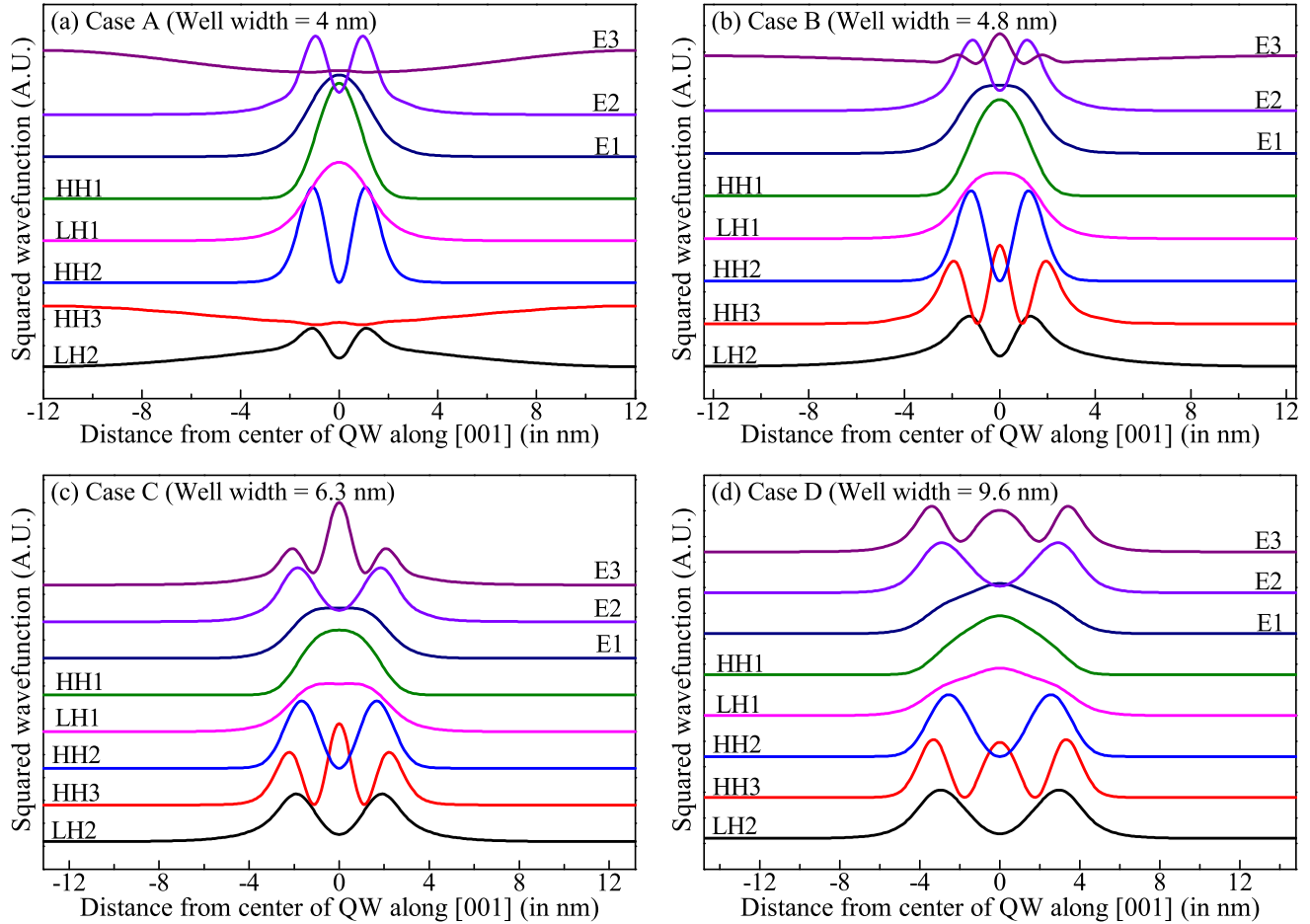}\label{}}
	\caption{First 3 electron and first 5 hole squared wavefunctions at \textbf{k} = 0 point for GaNBiAs QWs of (a: \textit{Top Left}) Case A (well width = 4 nm); (b: \textit{Top Right}) Case B (well width = 4.8 nm); (c: \textit{Bottom Left}) Case C (well width = 6.3 nm) and (d: \textit{Bottom Right}) Case D (well width = 9.6 nm) vs. distance from the center plane of QW along the [001] direction. E stands for electron, HH for heavy hole levels and LH for light hole levels.}
	\label{fig:wav-fn-ALL}
\end{figure*}

Fig. \ref{fig:wav-fn-ALL} shows the squared wavefunction of the first 3 electron and the first 5 hole energy levels at \textbf{k} = 0 point for Case A to D GaNBiAs/GaAs QWs. We can see that as the thickness increases, the E and H wavefunctions spreads out, as expected due to larger well width. In all the cases, the wavefunctions of E1 has a reasonable overlap with that of HH1, which causes the E1-HH1 fundamental transition. Transitions among higher E and H states follow the optical transition rule and depend on the extent of probability of wavefunction overlap.

On the whole, as summarized in Table \ref{tab:comparison}, we find that for our intended 1.55 $\mu$m (0.8 eV) GaAs-based fiber optic laser diode applications, the 6.3 nm thick GaN$_3$Bi$_{5.17}$As$_{91.83}$ lattice matched quantum well (Case C), gives the best optical performance. This is in comparison with the other three cases. Further fine tuning may lead us to better design parameters. GaNBiAs/GaAs QWs are excellent candidates for 1.55 $\mu$m optical applications. But it is important to carefully consider the design parameters which determine the final optical performances.

Having concluded our findings above, we would now like to highlight couple of important research aspects relating to dilute-N-Bi QW lasers. Firstly, it is well known that for bulk material, the conduction-heavy hole-split off hole-heavy hole (CHSH) Auger recombination is significant. However, there exists contradictory viewpoints in the QW domain. For example, Hausser \textit{et al.} have experimentally shown that the Auger recombination in QWs is about 3 times smaller than that in bulk \cite{hausser90}. Other works have reported a reduction in Auger recombination by two orders of magnitude in QWs compared to bulk \cite{chiu82}. These works show that the Auger recombination in QWs is of very minor significance, owing to which we have chosen to simplify our calculations. However, there are some counter arguments in existing literature which state that Auger recombination is a significant problem for long wavelength QW semiconductor lasers \cite{phillips99,tomic03}. In this context, more conclusive work is elicited to obtain a comprehensive understanding of the Auger effects in GaNBiAs QWs. In order to fully understand the nonradiative recombination, we would also like to trigger the investigation of the other two types of Auger recombination in GaNBiAs QWs: conduction-heavy hole-conduction-conduction (CHCC) and conduction-heavy hole-light hole-heavy hole (CHLH). We know that the CHSH Auger recombination can be suppressed by controlling the Bi composition (ensuring that the spin-orbit splitting energy exceeds the effective bandgap \cite{sweeney13}), and more readily so in the presence of N. But this does not continue to hold in case of CHCC or CHLH Auger recombination, for which there is no significant evidence in literature. Secondly, the major challenge with GaNBiAs QW laser is that, it is difficult to grow high quality material even with advanced molecular beam epitaxy or metal organic chemical vapor deposition. Incorporating both N and Bi requires different growth optimizations. One issue is that adding N into GaAs degrades the material quality \cite{fan02}. On the other hand, Bi as surfactant may increase the efficiency of N incorporation in GaNAs by up to 60\% \cite{young05}. However, too high a Bi \% can also be detrimental to the material quality. The Volz group has demonstrated GaBiAs QW lasers \cite{ludewig13} with 2.2\% Bi having a threshold current density of 1.56 kA/cm\textsuperscript{2}.

\section{\label{sec:sum}Summary and Conclusion}

We have studied the electronic bandstructure and optical properties of lattice matched GaNBiAs/GaAs quaternary alloy quantum well (QW) based on the 16-band \textbf{\textit{k$\cdot$p}} model. The different QW cases studied have their thicknesses and N : Bi : As composition tuned in such a way to have their emission peak at 1.55 $\mu$m, aiming towards optical communication systems. We have obtained insights on how the well thickness and (N,Bi) compositions affects the energy dispersion curves and closeness of electron and hole energy levels. An increase in the injection carrier density increases the maximum gain and also the gain spectral width. In our study we see that while thinner wells have decreasing differential gain at lower density, thicker wells show the same at higher density. It is, thus, of significant importance to determine the critical well thickness for a steady optical performance. Among our cases studied, we have identified the 6.3 nm thick GaN$_3$Bi$_{5.17}$As$_{91.83}$ lattice matched QW to exhibit optimized optical performance for our intended 1.55 $\mu$m (0.8 eV) optoelectronic device/system applications. Therefore, it is recommended that the design parameters such as material composition and well width must be carefully considered from the device application point-of-view for an accurate optimization of the optoelectronic performance characteristics.

\begin{acknowledgments}
	W. J. Fan would like to acknowledge the support from MOE Tier 1 funding RG 182/14.
\end{acknowledgments}

%\nocite{*}
%\bibliography{JAP-refs}% Produces the bibliography via BibTeX.
%

\end{document}